\documentclass[prd,aps,superscriptaddress,twocolumn,floatfix,nofootinbib]{revtex4-1}
\pdfoutput=1

\usepackage{amsfonts}
\usepackage{amsmath}
\usepackage{amssymb}
\usepackage{bm}
\usepackage{dcolumn}
\usepackage{graphicx}   
\usepackage[latin1]{inputenc}
\usepackage{latexsym}
\usepackage{rotating}
\usepackage{hyperref}
\usepackage{graphicx}
\usepackage{color}

\newcommand\be{\begin{equation}}
\newcommand\ba{\begin{eqnarray}}
\newcommand\ee{\end{equation}}
\newcommand\ea{\end{eqnarray}}

\begin{document}

\title{Why the DESI Results Should Not Be A Surprise}

\author{Robert Brandenberger}
\email{rhb@physics.mcgill.ca}
\affiliation{Department of Physics, McGill University, Montr\'{e}al,
  QC, H3A 2T8, Canada}


\begin{abstract}

The recent DESI results provide increasing evidence that the density of dark energy is time-dependent. I will recall why, from the point of view of fundamental theory,, this result should not be surprising.

\end{abstract}

\maketitle

\section{Introduction} 
\label{sec:intro}

DESI has just published results from their Data Release 2 \cite{DESI1} which (combined with CMB data from Planck \cite{Planck} and ACT \cite{ACT}) indicate a conflict at a level of greater than 3 sigma with the canonical $\Lambda$CDM model.  When testing the DESI baryon acoustic oscillation results in the context of a model which contains a parametrization of the dark energy equation of state $w$ in terms of two parameters $w_0$ and $w_a$
\be
w(a) \, = \, w_0 + w_a(1 - a) \, ,
\ee
where $a$ is the cosmological scale factor of a spatially flat universe normalized to $a = 1$ at the present time, and we recall that $w$ is the ratio of pressure density to energy density,  then the equation of state $w_0 = -1$ and $w_a = 0$ of a cosmological constant is ruled out at a level of greater than 3 sigma.  The data favours a time-dependent equation of state with $w_0 > -1$ and $w_a < 0$ which corresponds to a dark energy density which decreases as a function of time.  Note that this result is obtained without making use of supernova data. This result is much stronger than the data from the initial DESI data release \cite{DESI0} which showed some evidence for $w_0 > -1$ and $w_a < 0$, but with lower significance, and only when including supernova data \footnote{In fact, the evidence cannot be seen when making use of different parametrizations of the dark energy equation of state \cite{Guillaume}..}  

Additional evidence for a time-dependent dark energy density can be obtained \cite{DESI2} when adding the sum of neutrino masses $\sum_{i = 1}^3 m_{\nu}^i$ as a free parameter to the data analysis. In the context of the canonical $\Lambda$CDM model Bayesian evidence points to an unphysical negative value for this mass sum, but when analyzed in the context of a time-dependent dark energy density described by the two parameters $w_0$ and $w_a$, the Bayesian evidence returns to favouring a physically sensible (i.e. positive) value of the sum of masses.

From the observational point of view, these DESI results indicate the need for a paradigm shift in our understanding of late time cosmology. In the next section I will argue that, from the point of view of fundamental theory,  we should not have expected dark energy to be a remnant cosmological constant due to conceptual problems with the resulting cosmological model.  In fact, I will argue that to describe dark energy, we need to go beyond the standard effective field theory treatment in which spacetime evolves according to General Relativity and matter is described by an effective quantum field theory.  

In the following we consider a homogeneous, isotropic and spatially flat universe described by the metric
\be
ds^2 \, = \, dt^2 - a(t)^2 d{\bf{x}}^2 \, ,
\ee
where $t$ is physical time,  ${\bf{x}}$ are the comoving spatial coordinates and $a(t)$ is the scale factor.  In the following, the Hubble length scale
\be
l_H(t) \, \equiv \, H^{-1}(t) \, ,
\ee
where $H(t)$ is the instantaneous expansion rate, will play a key role. On length scales smaller than the Hubble radius, fluctuation modes simply oscillate, while on super-Hubble scales they freeze out and squeeze, leading to classicalization of initial quantum fluctuations (see e.g. \cite{MFB} for a review of the theory of cosmological fluctuations, and \cite{RHBfluctrev} for an introductory discussion).

\section{Trans-Planckian Censorship Criterion and the DESI Results}
\label{section2}

In the standard cosmological paradigm,  spacetime obeys the Einstein field equations and matter is considered to be a set of quantum fields.  In the effective field theory framework, each such field $\varphi({\bf{x}}, t)$ is expanded in comoving Fourier modes
\be
\varphi({\bf{x}}, t) \, = \, \int d^3{\bf{k}} e^{i {\bf{k}} {\bf{x}}} \varphi_k(t) \, ,
\ee
and each mode is quantized as a harmonic oscillator. At the level of linearized equations, each Fourier mode evolves independently. Gravitational and matter nonlinearities generate couplings between different modes. In order to avoid the Planck ultraviolet catastrophy, we need to impose an ultraviolet cutoff: no Fourier modes with physical wavenumber greater than some cutoff scale (expected to be either the Planck or the string scale) exist. In this setup,  a {\bf unitarity problem} arises \cite{Weiss}: in order to maintain the cutoff at a fixed physical scale, we have to continuously create modes. The Hilbert space of states becomes time-dependent. This is a serious problem for fundamental physics and is sometime called the {\it trans-Planckian problem} \cite{Jerome}.
 
Given an expanding universe,  a necessary condition was proposed \cite{TCC1} to ensure that physics which we currently observe be (at linear order) shielded from this unitarity problem: wavelengths which have to be ``created'' after the initial time $t_i$ , i.e. which are trans-Planckian before the initial time,  should never be able to exit the Hubble radius and hence classicalize. This condition (the {\it Trans-Planckian Censorship Criterium} (TCC)) can be written as
\be \label{TCC}
\frac{a(t)}{a(t_i)} l_{pl} \, < \, H^{-1}(t) 
\ee
for all times $t > t_i$, where $l_{pl}$ is the Planck length. For more detailed justifications of the TCC see e.g. \cite{RHB-TCC}. Note that (\ref{TCC}) should be viewed as a necessary but not in itself sufficient condition for a cosmology which is viable from the point of view of fundamental theory.  
 
 In a non-accelerating expanding cosmology the Hubble radius grows faster than the scale factor,  and hence the TCC is automatically satisfied.  In an accelerating universe, on the other hand, the TCC criterion is very constraining. Applied to inflationary cosmology, it implies that canonical inflationary models described in the framework of effective field theory are inconsistent \cite{TCC2}, unless the energy scale of inflation is very low, in which case it would take fine-tuning to obtain a sufficiently large amplitude of cosmological perturbations (see also \cite{Blamart}).
 
The TCC has crucial implications for the dark energy mystery: if dark energy is a cosmological constant, then the Hubble radius converges to a constant which is approximatly equal to $H_0^{-1}$ at late times, where $H_0$ is the present day Hubble expansion rate. In this case,  comoving scales continuously exit the Hubble radius. After a time interval $\Delta t$ given by 
\be
\Delta t \, = \, H_0^{-1} {\rm{ln}}  (l_{pl}^{-1} H_0^{-1}) \, ,
\ee
 scales which exit the Hubble radius will have been trans-Planckian today and hence in the zone of ignorance.  Thus, if dark energy were a cosmological constant,  the non-unitarity of the effective field theory would rear its ugly head after this time interval $\Delta t$ which is called the ``Trans-Planckian Censorship time''. To avoid this problem, accelerated expansion has to end, and hence dark energy has to be time-dependent \cite{TCC1, RHB-TCC}.

Note that the TCC constrains all dark energy models in which accelerated expansion does not end at some point in the future.
 
 There are a couple of important remarks. Firstly,  the TCC condition applies to any model which is based on effective field theory, and thus it applies much more generally than only to string theory.  On the other hand, if one were able to construct a theory which is not based on effective field theory, then the TCC criterion would not necessarily apply.  Recent efforts to obtain an accelerating cosmology using non-perturbative ingredients in string theory \cite{Keshav} (see also \cite{Dvali}) obtain accelerating phases with time-dependent equations of state.   Also, starting from a non-perturbative definition of superstring theory, the BFSS matrix model \cite{BFSS}, one obtains an emergent spacetime and early universe cosmology which does not yield a cosmological constant \cite{BBL} (see also \cite{BV} for earlier work).  
 
In the context of superstring theory, there are other arguments which indicate that dark energy must be time-dependent. FIrst of all, there are no-go ``theorems'' which indicate that it is not possible to obtain a positive cosmological constant from string theory using standard effective field theory techniques \cite{nogo}. Secondly,  string theory severely restricts the set of effective theory models which can be consistently derived from striing theory. Most effective field theories which one can write down are in the ``swampland'',  leaving only small island behind which form the ``string theory landscape'' (see \cite{swamp} for an original article and \cite{swamprevs} for reviews). In particular, the ``Swampland De Sitter'' (dS) conjecture \cite{dS} sets a lower bound on the slope of the potential of a scalar field which might dominate the evolution of the late time universe. In particular, this rules out dark energy being a cosmological constant (since this would correspond to a scalar field potential with zero slope). In fact, even quintessence models of dark energy are constrained by the  dS conjecture \cite{Lavinia}.
 
\section{Discussion}  

I have argued that from the point of view of fundamental theory, we should not be suprised that dark energy is time-dependent.  At least in the context of effective field theory, a model with a bare cosmological constant as dark energy would be non-unitary.  Formally, the argument is based on the TCC.  The TCC, in turn, can also be argued for in analogy with Penrose's black hole cosmic censorship hypothesis \cite{RHB-TCC}.
 
Unfortunately,  at this point fundamental theory does not yet explain what dark energy is, only what it is not.  It may be interesting to revisit scenarios of oscillating dark energy such as the ``Everpresent Lambda'' \cite{Sorkin} model, or the idea \cite{BR} that an oscillating dark energy is a remnant of the screening of a bare cosmological constant via the back-reaction of long  wavelength cosmological perturbations.

\begin{acknowledgements}

I am grateful to Carlos Frenk for highlighting to me the main points of the DESI results, and to my collaborators, in particular A. Bedroya,  C. Vafa,  M. Loverde,  and H. Bernardo, M. Blamart, S. Brahma, V. Kamali, G. Payeur and S. Laliberte,  on whose work I am drawing.  My research is supported in part by funds from NSERC and from the Canada Research Chair program.    

\end{acknowledgements}


\end{document}